\documentclass{article}
\usepackage{spconf,amsmath,graphicx,comment,booktabs,makecell,amsfonts}

\title{Multichannel Voice Trigger Detection based on Transform-average-concatenate}
\name{Takuya Higuchi$^1$, Avamarie Brueggeman$^2$*\thanks{*Work performed at Apple.}, Masood Delfarah$^1$, Stephen Shum$^1$}
\address{$^1$Apple, USA\\
  $^2$The University of Texas at Dallas, USA}
\begin{document}
%\ninept
%
\maketitle
\begin{abstract}
Voice triggering (VT) enables users to activate their devices by just speaking a trigger phrase. A front-end system is typically used to perform speech enhancement and/or separation, and produces multiple enhanced and/or separated signals. Since conventional VT systems take only single-channel audio as input, channel selection is performed. A drawback of this approach is that unselected channels are discarded, even if the discarded channels could contain useful information for VT. In this work, we propose multichannel acoustic models for VT, where the multichannel output from the frond-end is fed directly into a VT model. We adopt a transform-average-concatenate (TAC) block and modify the TAC block by incorporating the channel from the conventional channel selection so that the model can attend to a target speaker when multiple speakers are present. The proposed approach achieves up to $30\%$ reduction in the false rejection rate compared to the baseline channel selection approach.
%Furthermore, we reduce a computational complexity by only feeding averaged TAC output to the Transformer layer, which achieves a $??\%$ relative reduction in run time cost compared to a conventional TAC models while achieving a $??\%$ reduction in the false rejection rate compared to the baseline channel selection approach. 
\end{abstract}
\vspace{-0.1cm}
\section{Introduction}
\label{sec:intro}
\vspace{-0.2cm}
Voice trigger detection is an essential task for a voice assistant system, allowing a user to activate the voice assistant by simply speaking a wake word. Noise robustness is an important aspect of successful voice triggering (VT). A front-end speech enhancement and/or separation system is commonly used to improve the noise robustness \cite{8887564}. 
Speech separation is especially useful for VT and other downstream tasks when multiple speakers are present in a recording because a typical acoustic model cannot deal with speech mixtures. 

However, multiple separated and enhanced signals from the front-end system cannot directly be input to a typical VT system because the VT system takes only a single channel input. A simple solution to address this is channel selection, where one channel is selected from the multiple channels before VT \cite{HomePodFrontEnd}. A downside of this approach is that unselected channels are discarded even though they might contain useful information for VT. For example, if the front-end system wrongly suppresses a part of a target speech and introduces distortions in the selected channel, the suppressed part of the target speech could be contained in the other channels. 

We present a novel multichannel acoustic model for VT, where the multiple separated/enhanced channels from the front-end system are directly fed into a VT acoustic model. We adopt a recently-proposed transform-average-concatenate (TAC) block \cite{luo2020end} to perform inter channel processing within the acoustic model. In addition, we combine the selected channel in the TAC block so that the model is informed of the channel of most interest for VT. We conduct experimental evaluations on a far-field VT task, where the proposed multichannel approach outperforms a single channel baseline VT with channel selection by up to $30\%$ relative in terms of the false rejection rate (FRR). 
%Contributions of this paper are: 1) Adopt the TAC block for a VT task and compare with a conventional channel selection approach, and 2) Propose a new network architecture combining the selected channel with the TAC block. 
%selected channel in Explores new architectures combining the TAC and self-attention layers which can reduce run-time cost by up to $\%$ compared to the conventional TAC model.

%The remainder of the paper is organized as follows. Section \ref{sec:prior} describes the relation to prior work and contributions of the paper. Section \ref{sec:baseline} describes a baseline single channel VT system with channel selection, and then Section \ref{sec:proposed} presents our proposed approach. Section \ref{sec:exp} shows experimental results. Finally, Section \ref{sec:conc} concludes the paper.

\vspace{-0.1cm}
\section{Related work}
\label{sec:prior}
\vspace{-0.2cm}
Multichannel acoustic modeling has been investigated for far-field automatic speech recognition \cite{7859320}. Sainath et al. proposed using convolutional neural networks (CNNs) on multichannel time domain signals \cite{7404770,sainath2016factored} to directly learn both spatial and spectral characteristics from training data. A similar approach has also been used for keyword spotting \cite{ji2021end}.
%This end-to-end modeling has also been formulated in the frequency domain for efficient implementation \cite{sainath2016reducing}. Frequency domain multichannel modeling has also been explored in \cite{minhua2019frequency,kumatani2019multi}.
 More recently, attention based approaches have been proposed for multichannel acoustic modeling \cite{gong2021self, chang2021end}, where cross-channel attention is performed to learn from inter-channel characteristics. Although these approaches are end-to-end optimized for the target tasks, the model complexity and compute cost usually increase due to the joint spatial and spectral modeling, which is unsuitable for on-device applications such as VT. Moreover, there is no mechanism to differentiate a target speaker from interference speakers in these approaches.
%Although these approaches are end-to-end optimization, the model usually becomes microphone array dependent since the multichannel filter size of the CNN is determined by the number of microphones used during training. Moreover, these approaches increase model complexity as they model both spatial and spectral characteristics with the neural networks, which is unsuitable for on-device applications such as VT.

Other types of multichannel approaches have also been proposed for keyword spotting, where multichannel features are concatenated \cite{wu20_odyssey} or attention based weighted sum is performed on the multichannel features \cite{9054538,yu2020end}. Although these operations are simple and computationally light, they may not be enough to model inter-channel characteristics. 

Multichannel modeling has also been explored for neural network-based speech enhancement and separation. A TAC block is proposed for simple yet effective multichannel modeling for speech enhancement and separation \cite{luo2020end,taherian2022one,yoshioka2022vararray}. The TAC block is defined with simple channel-wise transformations, pooling and concatenation operations.

Our proposed multichannel VT modeling is based on the TAC block because
%it does not introduce an additional model complexity for spectral modeling and
it employs simple and light-weight operations for non-linear inter-channel modeling.
%This is especially important for the VT system that should be run on device.
%Moreover, unlike the prior multichannel modeling work \cite{7859320,luo2020end,yoshioka2022vararray} that uses features from raw multi-microphone recordings as the input, we use the output of the front-end system, i.e., enhanced and separated signals, as the multichannel input for the VT acoustic model following the prior work \cite{9054538,yu2020end}. This allows us to exploit the existing front-end system and use potentially more informative signals for VT than the raw multi-microphone signals. In addition, compared to \cite{9054538,yu2020end}, we incorporate channel selection to let the model focus on the target speaker (See Section \ref{sec:TACSC}). 
 For the multichannel input, we use the output of the front-end system, i.e., enhanced and separated signals similarly to the prior work \cite{9054538,yu2020end}. This allows us to exploit the existing front-end system and use potentially more informative signals for VT than the raw multi-microphone signals, while the model performs non-linear inter-channel operations with the TAC block in contrast to \cite{9054538,yu2020end}. Moreover, we incorporate channel selection to allow the model to focus on the target speaker (See Section \ref{sec:TACSC}).

\vspace{-0.1cm}
\section{Baseline system}
\label{sec:baseline}
\vspace{-0.2cm}
\begin{figure}[t]

\centerline{\includegraphics[width=7cm]{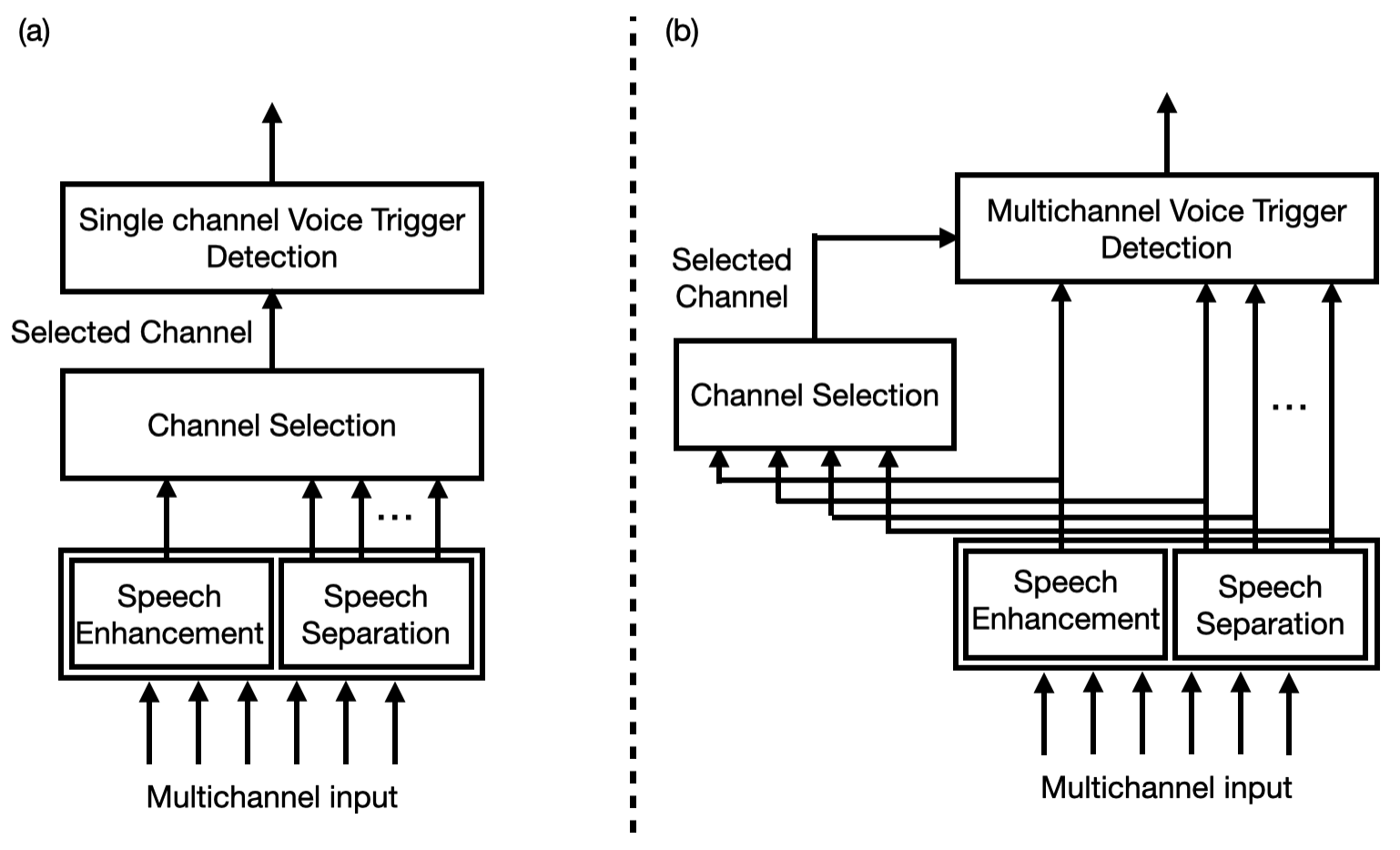}}
%  \vspace{2.0cm}
%
%\caption{(a) A conventional single channel voice trigger detection with a front-end speech enhancement and separation followed by channel selection. (b) The proposed multichannel voice trigger detection that takes the enhanced and separated speech signals as well as the selected channel as a multichannel input.}
\caption{(a) A conventional single channel voice trigger detection. (b) The proposed multichannel voice trigger detection with channel selection.}
\label{fig:mod_tac}
\vspace{-0.25cm}
%\vspace{-0.5cm}
%
\end{figure}
Figure \ref{fig:mod_tac} (a) shows the flowchart of a baseline system \cite{HomePodFrontEnd}. A front-end system consists of a speech enhancement module and a speech separation module. The enhancement module produces a single-channel enhanced speech signal, whereas the separation module produces $(N-1)$-channel output for $N-1$ separated signals. The separation module is especially useful when observed signals contain multiple speech signals, such as target speaker and TV noise. Then $N$ signals from both modules are fed into a VT system.

%VT models usually take a single channel input, so channel selection is required. 
The VT system employs a two-stage approach \cite{gruenstein2017cascade, HeySiri, sigtia2018efficient} to save run-time cost on-device. A small VT model (1st pass model) is always-on and takes streaming audio of each channel from the front-end. Once we detect an audio segment with a VT score exceeding a certain threshold, the audio segment is fed to a larger VT model and re-examined to reduce false activations. Since the VT models take only a single-channel audio, channel selection is performed using the 1st pass model \cite{HomePodFrontEnd}.

The 1st pass model processes each channel from the front-end system, and produces a VT score independently. Then, the channel with the highest 1st pass VT score is selected and fed into the 2nd pass model that has more complexity and is more accurate than the 1st pass model. This approach has an advantage when multiple speakers are present, because one channel containing a keyword phrase can be selected among multiple separated speech signals. However, a drawback of this approach is that unselected channels are discarded and not used for the 2nd pass model, whereas noise and interference signals in the discarded channels could also be useful for accurate VT in the 2nd stage. In addition, the speech enhancement or separation module could suppress the target speech and introduce distortions when there is no interference speech and/or background noise, which can be ineffective for VT. 

\vspace{-0.1cm}
\section{Proposed multichannel VT modeling}
\label{sec:proposed}
\vspace{-0.2cm}

In this paper, we propose multichannel acoustic models that can take the multichannel output from the front-end system. Figure \ref{fig:mod_tac} (b) shows the flowchart of our proposed system. While the 1st pass model still performs VT on each channel separately, the proposed 2nd pass acoustic model takes all the channels. In addition, the selected channel obtained with the conventional channel selection is also fed to the model in order to keep the advantage of the conventional approach on speech mixtures. We adopt and modify a recently-proposed TAC block for combining the multiple channels in a VT acoustic model.

\vspace{-0.05cm}
\subsection{Transform-average-concatenate (TAC)}
\label{sec:tac}
\vspace{-0.1cm}
Let us consider $N$ channel signals from the front-end system that performs speech enhancement and separation. Let $\mathbf{Z}_{i} \in \mathbb{R}^{T \times F}$ denote a time series of a $F$-dimensional feature from channel $i$. We first apply a linear layer and the parametric rectified linear unit (PReLU) activation function \cite{he2015delving} to $\mathbf{z}_{i,t}$:
\begin{align}
\mathbf{h}_{i,t} = P(\mathbf{z}_{i,t}), \label{eq:t}
\end{align}
where $\mathbf{z}_{i,t}$ denotes the feature vector at time $t$ for channel $i$ and $P(\cdot)$ denotes the linear transformation followed by the PReLU activation. Then, $\mathbf{h}_{i,t}$ is averaged across the channels and fed into another linear layer with the PReLU activation function as:
\begin{align}
\mathbf{h}^{avg}_{t} = Q(\frac{1}{N}\sum_{i}\mathbf{h}_{i,t}).\label{eq:a}
\end{align}
Then $\mathbf{h}^{avg}_{t}$ is concatenated with $\mathbf{h}_{i,t}$ and fed into a third linear layer and the PReLU activation function as:
\begin{align}
\hat{\mathbf{h}}_{i,t} = R([\mathbf{h}_{i,t};\mathbf{h}^{avg}_{t}]).\label{eq:c}
\end{align}
Finally, a residual connection is applied to obtain an output of the TAC block as:
\begin{align}
\hat{\mathbf{z}}_{i,t} = \mathbf{z}_{i,t} + \hat{\mathbf{h}}_{i,t}.\label{eq:res}
\end{align}
These operations enable learning inter-channel characteristics with the simple channel-wise transformations and the pooling operation. Note that all the operations in the TAC block is permutation invariant between the channels by design for microphone-array agnostic modeling, which allows us to feed the arbitrary number of separated/enhanced signals into the TAC block.

\begin{figure}[t]

\centerline{\includegraphics[width=4cm]{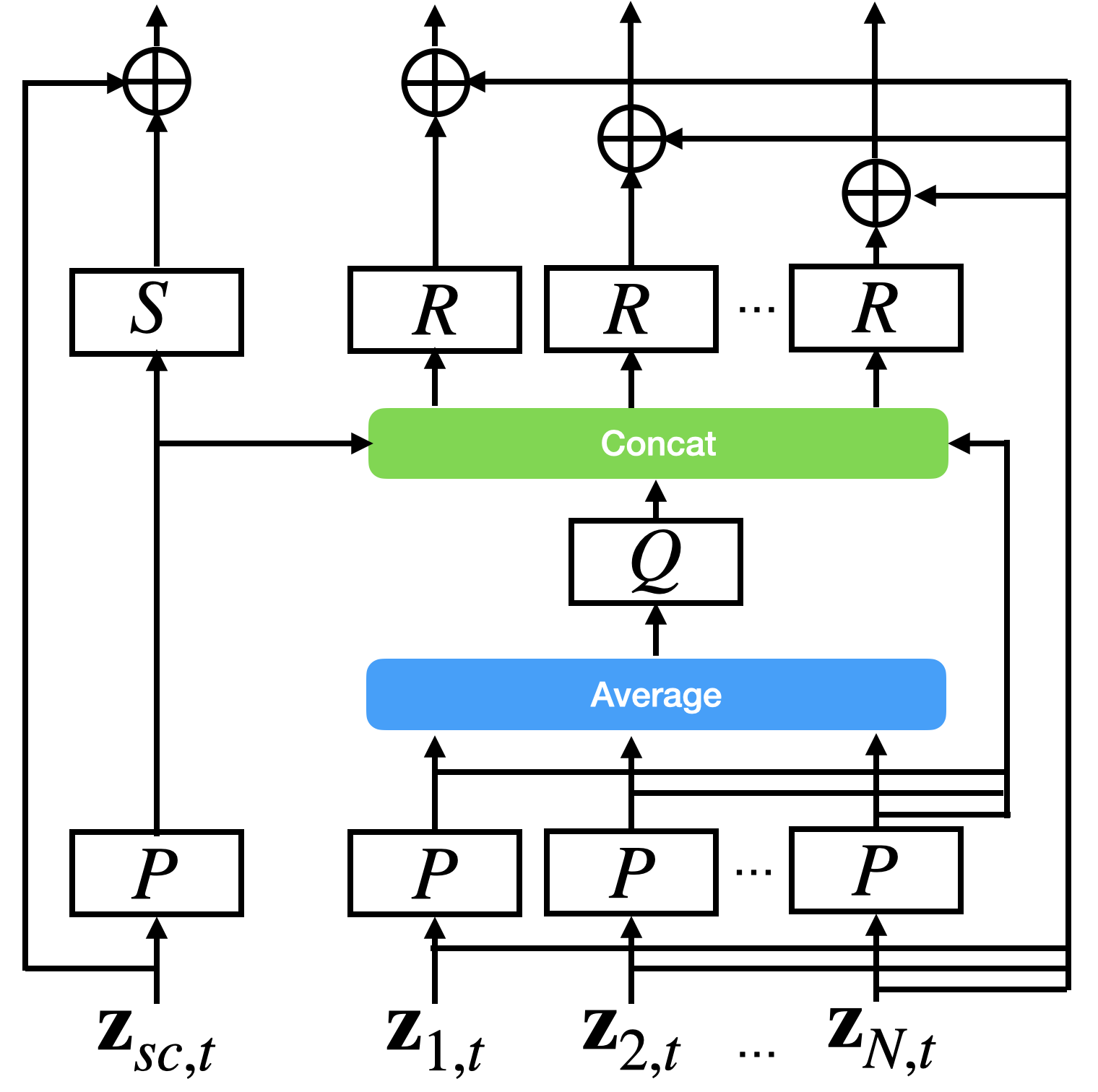}}
%  \vspace{2.0cm}
%
\caption{The modified TAC block with the selected channel.}
\label{fig:mod_tac}
\vspace{-0.25cm}
%\vspace{-0.5cm}
%
\end{figure}

\vspace{-0.05cm}
\subsection{Modified TAC with selected channel}
\label{sec:TACSC}
\vspace{-0.1cm}
Although the permutation-invariant operations enable mic-array-agnostic speech separation in the previous literature \cite{yoshioka2022vararray}, its permutation-invariant nature would be problematic when one of the channels is of more interest for VT. For example, the front-end system performs speech separation and produces multiple channels which contain either target or interference speech at each channel. A VT model should attend to the target speech to perform VT detection. However, the TAC block processes every channel equally, which confuses the VT model during training and inference when multiple speakers are present.

To address this, we propose exploiting the selected channel obtained with the conventional channel selection approach. Figure \ref{fig:mod_tac} shows the proposed block obtained by modifying the conventional TAC block. Let $\mathbf{z}_{sc,t}$ denote a feature vector of the selected channel. The modified TAC block takes a feature vector $\mathbf{z}_{i,t}$ for channel $i  (=1,...,N)$ as well as $\mathbf{z}_{sc,t}$. We first apply eq. (\ref{eq:t}) to $\mathbf{z}_{sc,t}$ as with $\mathbf{z}_{i,t}$:
\begin{align}
\mathbf{h}_{sc,t} = P(\mathbf{z}_{sc,t}), \label{eq:t_sc}
\end{align}
while the average operation is performed on $\mathbf{h}_{i,t} (i=1,...,N)$ using eq. (\ref{eq:a}). Then, $\mathbf{h}_{sc,t}$ is concatenated with  $\mathbf{h}_{i,t}$ and $\mathbf{h}^{avg}_{t}$, and fed into a linear layer and the PReLU activation function as:
\begin{align}
\hat{\mathbf{h}}_{i,t} = R([\mathbf{h}_{i,t};\mathbf{h}^{avg}_{t};\mathbf{h}_{sc,t}]).\label{eq:c_sc}
\end{align}
This operation distinguishes the selected channel from the other channels and encourages the model to learn from the selected channel. Finally, another linear layer and the PReLU activation function is applied to  $\mathbf{h}_{sc,t}$ to reduce the dimensionality to dimensionality of the input before the residual connection:
\begin{align}
\hat{\mathbf{z}}_{sc,t} = \mathbf{z}_{sc,t} + S(\mathbf{h}_{sc,t}),\label{eq:res_sc}
\end{align}
while $\hat{\mathbf{z}}_{i,t}$ $(i=1,...,N)$ is obtained with eq. (\ref{eq:c}). 

\vspace{-0.05cm}
\subsection{Acoustic modeling with self-attention layers}
\label{sec:ac}
\vspace{-0.1cm}
The TAC block was combined with self-attention layers in \cite{yoshioka2022vararray}, where self-attention is performed for each output channel of TAC blocks. This approach drastically increases a run-time computation cost because a quadratic self-attention operation is repeated for each channel, which is unsuitable for VT that should be run on-device with low latency. To alleviate this, we apply an average pooling layer before feeding a multichannel output from the TAC block to the self-attention layers for temporal and spectral modeling. 
% Then, the averaged single channel output is fed into the self-attention layers for temporal and spectral modeling for VT. 

%We explore two different approaches which reduce the compute cost from VarArray. A naive approach is to apply a pooling layer and reduce the number of channel before the self-attention layers (figure \ref{} (b)). This model first uses the stacked TAC layers to model inter-channel characteristics, then the pooling layer converts the multichannel output from the TAC layers to a single channel. The single channel output is then used by the self-attention layers. 

%Alternately, we can use the self-attention layers between the TAC layers, but feed only the averaged output to the self-attention layers (figure \ref{} (c)), which allows us to perform self-attention across time direction only once regardless of the number of input channels. Thus, this approach also prevents the quadratic computation on every channel and reduces the compute cost.

\vspace{-0.1cm}
\section{Experimental evaluation}
\label{sec:exp}
\vspace{-0.2cm}

We evaluated the effectiveness of the proposed approach on a far-field VT task. Since some practical use cases (e.g., the presence of playback/interference speakers) are not well-represented in common public datasets, we used our in-house dataset for evaluation. The proposed approach was compared with a conventional single channel VT that used channel selection. It should be noted that a simple concatenation of multichannel features used in \cite{wu20_odyssey} did not outperform the single channel baseline in our preliminary experiments.

\vspace{-0.05cm}
\subsection{Data}
\vspace{-0.1cm}
For training, we used $\sim2.3$ million human-transcribed single channel utterances. Multichannel reverberant signals were simulated by convolving measured room impulse responses (RIRs), which were recorded in various rooms and microphone locations with a six channel microphone-array. In addition, roughly $20\%$ of utterances were augmented by convolving simulated four-channel RIRs and then adding multichannel non-speech noise signals or multichannel playback. Then we combined these three types of utterances to obtain a simulated multichannel training dataset. Finally, the multichannel signals were fed into the front-end system to obtain one enhanced signal and three separated signals for each utterance.

For evaluation, we used an in-house dataset, where positive samples were collected in controlled conditions from 100 participants. Each participant spoke the keyword phrase to the smart speaker with six microphones. Note that there was a mismatch in microphone arrays used for a part of training data  and test data, which was compensated for by the front-end that always produced four channels. The recordings were made in various rooms in four different acoustic conditions: quiet (no playback, no noise), external noise, e.g., from TV or appliances, music playing from the device at medium volume, and music playing at loud volume. 1300 such positive samples were collected. For negative data, we collected 2000 hours of audio by playing podcasts, audiobooks, etc, which did not contain the keyword phrase. The negative samples were also recorded with the smart speaker. The same front-end system was applied to the evaluation dataset to obtain enhanced and separated signals for each sample.
\begin{comment}
\begin{figure*}[t!]
\begin{center}
\centerline{\includegraphics[width=12cm]{DETs_v3.png}}
%  \vspace{2.0cm}
\vspace{-0.25cm}
\caption{DET curves for the (a) quiet, (b) noisy, (c) medium and (d) loud volume music playback conditions.}
\label{fig:DET_all}
\vspace{-1.0cm}
\end{center}
\end{figure*}
\end{comment}

\vspace{-0.05cm}
\subsection{Settings}
\vspace{-0.1cm}
For the front-end, we used echo cancellation and dereverberation followed by a mask-based beamformer for speech enhancement and blind source separation. See \cite{HomePodFrontEnd} for more details of the front-end. The speech separation module produced three separated signals, and so we obtained four channel signals in total from the front-end. It should be noted that our proposed model architecture can be used with any front-end that produces a multichannel output. 

For the 1st pass VT and channel selection, we used 5x64 fully-connected deep neural networks (DNNs). The 5x64 DNNs predicted a frame-wise posterior for 20 classes: 18 phoneme classes for the keyword, one for silence and one for other speech. Then a hidden Markov model (HMM) decoder produced a VT score and alignment for a trigger phrase based on the posteriors in a streaming fashion. This 1st pass model was run on the four channels separately and produced four VT scores. Then, a trigger segment in the channel with the highest score was input to the larger VT model in the second stage for the baseline.

For the VT models in the second stage, we used a Transformer encoder \cite{vaswani2017attention,adya2020hybrid} as an acoustic model. The baseline single channel model consisted of six Transformer encoder blocks, each of which had a multi-head self-attention layer with 256 hidden units and 4 heads, followed by a feed-forward layer with 1024 hidden units. Finally a linear layer transformed the output from the Transformer blocks to logits for 54 phoneme labels and one blank label for a Connectionist Temporal Classification (CTC) loss \cite{graves2006connectionist}. A VT score was obtained by computing a decoding score for the wake word. The baseline model used 40-dimensional log-mel filter bank features with $\pm$ 3 context frames as input. 

For the proposed multichannel model, we simply prepended one original/modified TAC block and an average pooling layer to the baseline model. The modified TAC block had $3\times256$ hidden units for $P$ and $Q$, and $280$ units for $R$ and $S$. 
%Also, the all operations in the modified TAC block are frame-wise, and so it has much less compute compared to the self-attention layers which requires a quadratic computation in the time dimension.
%Since we used enhanced and separated signals from the front-end as input, we used only log-mel filter bank features from the four channels in contrast to prior work \cite{yoshioka2022vararray} that also used phase difference features.
We used the log-mel filter bank features from the four channels as input. 
%The multichannel model used the four channels from the front-end and processed them as described in Section \ref{sec:proposed}.
Since our training data was not a keyword specific dataset on which the channel selection could be performed, we used the channel from the speech enhancement module as a pseudo selected channel during training. We also compared with a standard TAC block where the modified TAC block with the selected channel was replaced with the TAC that took only the four channels without knowing which one was the selected channel. 
The numbers of model parameters for the baseline, the proposed models with the standard and modified TAC were $4.8M$, $6.1M$ and $6.5M$, respectively. It should be noted that simply increasing the model size of the baseline single channel model did not improve a VT performance in our preliminary experiments.

All the models were trained with the CTC loss using the Adam optimizer \cite{kingma2014adam}. The learning rate was initially set at $0.0005$, then gradually decreased by a factor of $4$ after $10$th epoch, until we finished training after $28$ epochs. We used $32$ GPUs for each model training and the batch size was set at $128$ at each GPU.

\vspace{-0.2cm}
\subsection{Results}
\vspace{-0.2cm}
\begin{comment}
\begin{figure}[t]

\centerline{\includegraphics[width=7cm]{quiet_det.curve.Quasar_checker.txt.png}}
%  \vspace{2.0cm}
%
\caption{DET curves for the quiet condition.}
\label{fig:DET_quiet}
\vspace{-0.25cm}
%
\end{figure}
\begin{figure}[t]

\centerline{\includegraphics[width=7cm]{noise_det.curve.Quasar_checker.txt.png}}
%  \vspace{2.0cm}
%
\caption{DET curves for the noisy condition.}
\label{fig:DET_noise}
\vspace{-0.25cm}
%
\end{figure}
\begin{figure}[t]

\centerline{\includegraphics[width=7cm]{bargeInMusic_Medium_det.curve.Quasar_checker.txt.png}}
%  \vspace{2.0cm}
%
\caption{DET curves for the medium volume playback condition.}
\label{fig:DET_medium}
\vspace{-0.25cm}
%
\end{figure}
\begin{figure}[t]

\centerline{\includegraphics[width=7cm]{bargeInMusic_Loud_det.curve.Quasar_checker.txt.png}}
%  \vspace{2.0cm}
%
\caption{DET curves for the loud volume playback condition.}
\label{fig:DET_loud}
\vspace{-0.25cm}
%
\end{figure}
\end{comment}

\begin{table}[t]
\vspace{-3mm}
  \caption{False rejection rates [$\%$] for different conditions at an operating point of 1 FA/100 hrs.}
  %\vspace{-3mm}
  \label{tab:FRRs}
   \centering
   \scalebox{0.75}{
\begin{tabular}{cccccc}
  \toprule
 &  quiet & noisy & \makecell{medium\\ playback} & \makecell{loud\\ playback} & overall\\
  \midrule
 baseline & 4.48 &\textbf{6.75}& 5.23 & 15.32 & 7.74\\ \midrule
 proposed & \textbf{2.25} &10.7& \textbf{3.78} & \textbf{12.46} & 7.20\\ \midrule
  \makecell{proposed\\ (+ selected channel)} & 3.12 &7.12& 4.55 & 14.62 & \textbf{7.16}\\ \bottomrule
\end{tabular}
}
\vspace{-0.5cm}
\end{table}

Table \ref{tab:FRRs} shows FRRs with a threshold that gives $0.01$ FA/hr on the overall dataset for each model. 
%Figure \ref{fig:DET_all} shows detection error trade-off (DET) curves for the four different conditions. 
%$proposed$ and $proposed+sc$ indicate our multichannel approach with the standard and modified TAC blocks, respectively.
In the quiet condition, the proposed multichannel models outperformed the baseline with a large margin. This could be because the speech enhancement and separation were unnecessary in this case and would introduce distortions to the target speech in the selected channel, while the proposed models compensated the distortions by looking at all four channels (plus the selected one). In the music playback conditions, we observed moderate improvements with the proposed models. This could be because the multichannel models could learn echo residuals more effectively from the multichannel signals, where different front-end processing was applied at each channel. In the noisy condition, the vanilla TAC regressed compared to the baseline. We found that failure cases contained speech interference from TV. This is reasonable because there is no cue for the vanilla TAC to determine the target speaker when multiple speakers are present in the separated signals. By incorporating the selected channel, the proposed approach achieved a similar performance on the noisy condition while outperforming the baseline on the other conditions.
%Table \ref{tab:FRRs} shows FRRs with a threshold that gives $0.01$ FA/hr on the overall dataset for each model. 
%Table \ref{tab:FRRs} shows FRRs at an operating point (=$0.01$ FA/hr) for the different conditions.
%The proposed multichannel model without the selected channel achieved the best performances on quiet, medium and loud playback conditions. However, a large regression was observed for the noisy condition.
%This was because the noisy condition included speech interference from TV, which confused the model because the model did not have any cue to determine which one was the target speaker. On the other hand, by incorporating the selected channel, the proposed approach was able to keep a comparable performance on the noisy condition while achieving improvements over the baseline for the other conditions.
The proposed approach with the selected channel achieved $30\%$, $13\%$ and $4.6\%$ relative reductions in FRRs on the quiet, medium and loud volume playback conditions, respectively, and a $7.5\%$ FRR reduction on the overall dataset compared to the single channel baseline. These results show the effectiveness of the proposed approaches.

\vspace{-0.2cm}
\section{Conclusions}
\label{sec:conc}
\vspace{-0.4cm}
In this paper, we propose multichannel acoustic modeling for VT based on the TAC block. The multichannel acoustic model directly takes multichannel enhanced and separated signals from the front-end and produces a VT score. We further modify the original TAC block by incorporating the selected channel to deal with speech mixtures. The experimental results show that the proposed multichannel model outperforms the single channel baseline in the quiet and playback conditions, and achieves a similar performance in the noisy condition.
%Our future work includes exploring other front-end systems for the proposed approach as well as exploiting raw microphone signals besides the processed signals. 

\vspace{-0.2cm}
\section{Acknowledgement}
\label{sec:ack}
\vspace{-0.4cm}
We thank Mehrez Souden for his feedback on the paper and the helpful discussions.

\vfill\pagebreak

\bibliographystyle{IEEEbib_short}
\bibliography{mybib}

\end{document}